\documentclass[conference]{IEEEtran}
\ifCLASSINFOpdf
\usepackage[pdftex]{graphicx}
\usepackage{todonotes}
\usepackage{url}

\usepackage{color, colortbl}
\else
\fi
\begin{document}
%
\title{Video Quality Enhancement Using Deep Learning-Based Prediction Models for Quantized DCT Coefficients in MPEG I-frames}


\author{\IEEEauthorblockN{Antonio J. G. Busson, Paulo R. C. Mendes, Daniel de S. Moraes,\\ Álvaro M. da Veiga, Álan L. V. Guedes, and Sérgio Colcher}
\IEEEauthorblockA{TeleMidia Lab - Department of Informatics\\Pontifical Catholic University of Rio de Janeiro\\\{busson, daniel, paulo.mendes, alvaro, alan\}@telemidia.puc-rio.br, colcher@inf.puc-rio.br}}

%


\maketitle

\begin{abstract}
Recent works have successfully applied some types of Convolutional Neural Networks (CNNs) to reduce the noticeable distortion resulting from the lossy JPEG/MPEG compression technique. Most of them are built upon the processing made on the spatial domain. In this work, we propose a MPEG video decoder that is purely based on the frequency-to-frequency domain: it reads the quantized DCT coefficients received from a low-quality I-frames bitstream and, using a deep learning-based model, predicts the missing coefficients in order to recompose the same frames with enhanced quality. In experiments with a video dataset, our best model was able to improve from frames with quantized DCT coefficients corresponding to a Quality Factor (QF) of 10 to enhanced quality frames with QF slightly near to 20.
\end{abstract}

\begin{IEEEkeywords}
Deep learning, Convolutional Neural Networks, Video compression, MPEG, DCT.
\end{IEEEkeywords}

%
\IEEEpeerreviewmaketitle

\section{Introduction}
\label{sec:intro}

The application of methods based on Deep Learning (DL) in multimedia systems has opened a range of cognitive features in many directions that go beyond the traditional functionalities of capturing, streaming and presenting information. It has provided a whole new extent of capabilities that includes detection and classification of objects. New platforms and development techniques were tailored, and entirely new frameworks were brought together to enhance the development of such systems~\cite{Busson2019-webmedia} trying to fill in the gap between this vast (and relatively new) technological knowledge and the practical development of modern systems.

But the impact of DL techniques did not stay confined to these new semantic functionalities; representation, coding and compression of information from different media types have also been impacted by the advances brought by DL. In fact, the application of Machine Learning (ML) techniques to enhance the quality and performance of image coding and compression has appeared even before the recent reemergence of Neural Networks (NN) or DL architectures. Caputo~\cite{caputo_webmedia}, for instance, used genetic algorithms to optimize parameters for a video compression task in early 2008.

Many recent works have successfully applied DL architectures, specially some types of Convolutional Neural Networks (CNNs), to reduce the noticeable distortion resulting from the lossy JPEG/MPEG compression technique. Many of them are built upon the processing made entirely on the spatial domain, usually based on learned filters applied directly over the image pixels or blocks~\cite{zhang2017beyond,dong2015image,arcnn2015,farcnn2016}, while others opt to make adjustments (and learning) in the frequency domain (usually combined with spatial information) based on the Discrete Cosine Transform (DCT) which is part of the compression process~\cite{rajesh2019dctcompcnn, verma2018dct, zhang2018dmcnn, sfcnn}.

In this work, we propose an MPEG video decoder that incorporates a DL model trained to recover lost coefficients in the MPEG-encoding quantization process. Unlike other works, our method is purely based on the frequency-to-frequency domain, built upon the received quantized DCT coefficients from a bitstream of lower-quality \textit{I-frames}. The decoder predicts the missing coefficients and recomposes the corresponding \textit{I-frames} using information from a higher quality quantization table. This approach has advantages over methods based only on the spatial domain since it is possible to apply it to recover DCT coefficients directly in the media bitstream before its decoding. MPEG based systems can use this method to store and stream videos in low quality, and display them in a higher quality using our decoder without any changes to the MPEG compression algorithm, which means less disk space and bandwidth usage.

We evaluate how CNNs architectures that were initially designed for the spatial-to-spatial domain perform in the frequency-to-frequency domain. So, instead of using pixels from the frames to feed CNNs, we use low-quality quantized DCT coefficients with a Quality Factor of 10 (QF = 10) as input, and higher quality DCT coefficients (QF = 50) as a reference for learning. In addition, we also tested DCT coefficients with QF = 30 and QF = 40 to test whether the models converged better than with QF = 50.

This paper is organized as follows. We begin, in Section~\ref{sec:jpeg}, by briefly summarizing the essential aspects involved in coding and decoding images using the MPEG standard. Then, in Section~\ref{sec:related}, we focus on how some recent related works have been successfully applying CNNs and DL architectures in order to increase image quality and performance over the MPEG compression process. In Section~\ref{sec:model} we introduce our proposal for a decoder that incorporates the DL model to recover lost coefficients from the MPEG encoding quantization process, followed by Section~\ref{sec:experiment} where we describe the experiments conducted to evaluate the effectiveness of our proposal. Section~\ref{sec:conclusion} is devoted to our final remarks and conclusions.
\section{MPEG-Compression}
\label{sec:jpeg}

MPEG video coding standards family (ISO/IEC JTC 1/SC 29)\footnote{\url{https://www.iso.org/committee/45316.html}} is a popular method of lossy compression for video/audio. 
It is extremely portable and compatible with almost any video/audio processing applications, such as digital media storage~(e.g. Blue-ray), web-based video consumption~(e.g. mobile device, PC), dedicated video communication hardware~(e.g. video conference, hearth care), and video broadcast (e.g. TV systems).
MPEG provides a highly controlled degree of compression, allowing a selectable tradeoff between storage size and perceptual quality. 
Based on a psycho-visual model, it reduces or completely discards information in certain frequencies and areas of the picture that the human eye has limited ability to fully perceive. It also exploits temporal (over time) and spatial (across frames) redundancy common in video to achieve better data compression.

Some of the main concepts of the MPEG video coding are proposed in the MPEG-1 format, which has three frame class: (1) \textit{I-frame} (or keyframe), which contains all the information and does not depend on the content of any previously decoded frame; (2) \textit{P-frame}, which store only the difference in an image from the frame (either an \textit{I-frame} or \textit{P-frame}) immediately preceding it. This frame class exists to improve compression by exploiting the temporal (over time) redundancy in a video; And (3) \textit{B-frame}, which is referred to as bi-directional interpolated prediction frames, rely on the frames preceding and following them. 

\begin{figure}[!ht]
    \centering
    \includegraphics[width=\linewidth]{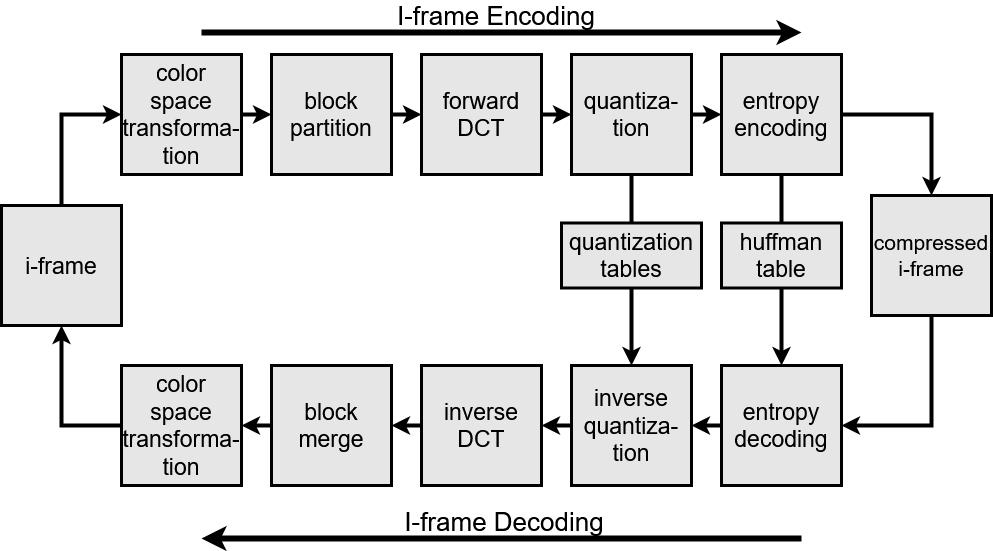}
    \caption{I-frame Codec.}
    \label{fig:i_frame}
\end{figure}

In this work, we are focused on increasing the video quality by extending specifically the  MPEG \textit{I-frame}, since it contains most of the relevant information. Fig.~\ref{fig:i_frame} illustrates the \textit{I-frame} codec process. In the remainder of this section, we describe in detail how it works.

The MPEG \textit{I-frame} encoding is identical to the JPEG encoding~\cite{joint2004jpeg}. It consists of five main stages:
\begin{description}
    \item {\textbf{Color space transformation:}} the image is converted from \textit{RGB} to \textit{YCbCr}, where \textit{Y} channel represents the brightness, \textit{Cb} and \textit{Cr} channels represents the chrominance of blue and red components;
    \item {\textbf{8x8 block partition:}} each channel is partitioned into non overlapping blocks of size 8x8;
    \item {\textbf{Forward DCT:}} each 8x8 block of each channel is level-shifted (i.e., each pixel value is subtracted by 128) to produce a data range that is centered on zero and then converted to a frequency-domain representation, using a normalized, two-dimensional type-II discrete cosine transform (DCT);
    \item {\textbf{Quantization:}} each DCT coefficient of each 8x8 block is divided by the corresponding quantization table and rounded to an integer (this rounding operation is the lossy operation in the whole process);
    \item {\textbf{Entropy coding:}} the frequencies components of each 8x8 block are transformed in a 1x64 vector. As this scan is done in the zig-zag method, this results in a vector ordered from low frequencies to high frequencies coefficients values. Since the previous quantization step reduced many of theses high-frequency values to zero, there is usually a long run of zero coefficients to the end of the vector. The \textit{Run-Length Encoding} (RLE) algorithm is then applied, transforming some long sequence of zeros in a single pair of numbers (code, number of occurrences). After that, Huffman changes the encoding, using fewer bits to encode frequently-used symbols and more bits to rarely-used ones.
\end{description}

The decoding phase reverses these steps, which involves entropy decoding, inverse quantization, inverse DCT (IDCT), level shifting, 8x8 block merging, and color space conversion from \textit{YCbCr} to \textit{RGB}.

\begin{figure}[!ht]
    \centering
    \includegraphics[width=0.45\textwidth]{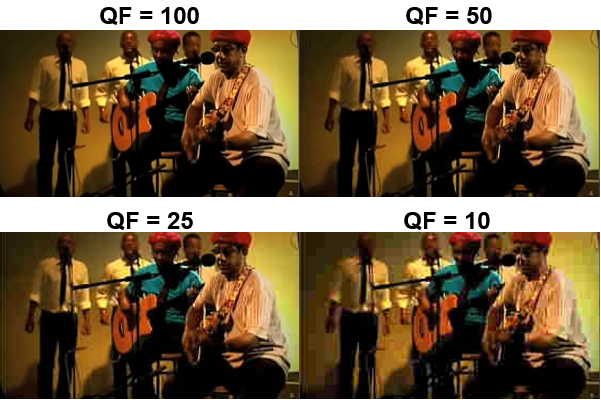}
    \caption{Frames degradation corresponding to their respective Quality Factor (QF) used in MPEG-Compression.}
    \label{fig:jpeg_deagratation}
\end{figure}

The use of the \textit{YCbCr} color space allows for greater compression without a significant effect on the perceptual quality of the image. Brightness information is more important than chrominance for the human visual system. That is why MPEG uses two tables in the quantization stage. The table used to quantify the brightness (\textit{Y}) channel preserves more information than the table used to quantify the chrominance channels (\textit{CbCr}). Additionally, MPEG allows the application to configure the Quality Factor (QF) used for compression. The quality setting operates in a range of 0-100 and is used to scale the values of the standardized quantization table. Fig.~\ref{fig:jpeg_deagratation} shows an example of frame degradation according to the different QFs used in MPEG-Compression.

It can be noticed from Fig.~\ref{fig:jpeg_deagratation} that the decoded image with QF=50 is already showing some initial signs of compression artifacts,\footnote{A compression artifact is a noticeable distortion caused by the application of a lossy compression technique; in this specific case of the MPEG I-frames, the main source of compression artifacts is the quantization of DCT blocks that lies at the heart of the compression standard algorithm.}  but still remains visually similar to the decoded image with QF=100 (the highest quality). As the QF value decreases, the resulting image shows increasing levels of distortions or degradation, such as loss of sharpness, loss of color detail, and blocking structure.
\section{Related Work}
\label{sec:related}

Many works have been devoted to the problem of reducing compression artifacts of images and videos. Many of them have focused on the enhancement of JPEG compressed images focusing on the processing made entirely on the spatial domain, applying filters to deal with specific aspects of the image, such as edge and texture. Others use the DCT domain information combined with spatial information to accomplish the same task. As the JPEG and MPEG I-frames compression/decompression pipelines are very similar, these works are closely related to ours.

For instance, the DnCNN~\cite{zhang2017beyond} is one of the most prominent architectures that operates in spatial domain. It is based on a feed-forward CNN that is commonly applied to image denoising, JPEG-deblocking and image super-resolution tasks.
Dong \textit{et. al}~\cite{dong2015image} follow a similar path with a model that aims at learning an end-to-end mapping between low and high-resolution images called SRCNN (Super-Resolution Convolutional Neural Networks). 
Their model is composed of three convolutional layers, each responsible for a specific task: (1) extraction of overlapping patches from the input and their representation as high-dimensional vectors; (2) non-linear mapping between each high-dimensional vector to another high-dimensional one; and (3) aggregation of these patch-wise representations to generate the final output. The model achieves good performance on the problem of single image super-resolution. 

As an improvement to the SRCNN, the same authors proposed the AR-CNN (Artifacts Reduction Convolutional Neural Networks) \cite{arcnn2015} and the Fast-ARCNN~\cite{farcnn2016}, aiming at reducing compression artifacts with the addition of one or more layers to clean noisy features. They reuse the features learned in a shallow network, transferring them to a deeper architecture with specific fine-tuning techniques. Results show the effectiveness of the approach in suppressing blocking artifacts and in maintaining edge patterns and sharp details. The Fast-ARCNN differs from the original model by having a ``shrinking'' layer between the first two layers, which speed up the inference process, and by using large-stride convolution filters in the first layer (and the corresponding deconvolution filters in the last layer). According to the authors, the Fast AR-CNN, can be 7.5 times faster than the original AR-CNN with almost no loss of quality.


There are works that use images represented in the DCT domain as the input for image classifiers.
Rajesh \textit{et. al}~\cite{rajesh2019dctcompcnn} propose the DCT-CompCNN, a CNN that uses DCT coefficients as the input for a classification task. They tested this modified input representation by comparing the existing ResNet-50\cite{resnet} architecture and their proposed architecture using a public dataset, reporting a better performance.

Verma \textit{et. al} ~\cite{verma2018dct} use a CNN in the DCT domain to verify the authenticity of an image by determining the number of times it has been passed through lossy compression algorithms. First the image is divided into blocks and their DCT coefficients are calculated. Then, they compute the histogram of the frequencies present in each block and use it as the input to the CNN. Finally, with the result obtained for each block, they perform a voting to produce the result for the entire image.


Additionally, other works have acknowledged that the DCT coefficients might be helpful in the task of restoring JPEG compressed images, and have used features of images in both domains, spatial and DCT, as input for deep learning models. 
Zhang \textit{et. al}~\cite{zhang2018dmcnn} propose the DMCNN, which is a CNN that uses both DCT and spatial domain information for the removal of artifacts generated by the JPEG lossy compression. The author's argument that the redundancies present in both the spatial and DCT domain may be used to improve the performance of the model. 
In the same direction, Kim \textit{et. al}~\cite{sfcnn} propose SF-CNN, which also work in both spatial and DCT domains. In their architecture, each 8x8 block in the DCT domain is transformed in a ``pixel'' with 64 channels, where each channel represents a DCT frequency. They demonstrate that the use of DCT coefficients may be useful for restoring JPEG compressed images.

Regarding the enhancement of compressed video quality, there are works such as Yang \emph{et. al}~\cite{yanghevc} that proposes a Decoder-side Scalable Convolutional Neural Network~(DS-CNN) approach to achieve quality enhancement for the High Efficiency Video Coding (HEVC). Their approach can handle both intra and inter coding distortions by enhancing I and B/P frames of HEVC.

In another work, Yang \textit{et. al}~\cite{yang2020learning} use a Support Vector Machine (SVM) to detect high quality frames, what they call Peak Quality Frames (PQF).  
Then, an architecture named Multi-Frame CNN (MF-CNN) is proposed for quality enhancement, in which both the current frame and its adjacent PQFs are the inputs. Nevertheless, both works approach the task using the spatial domain.

Unlike these previous works, we intend to explore deep learning models to restore MPEG I-frames based on the DCT coefficients alone (without any spatial domain information). These works also perform the restoration just of channel \textit{Y} or perform the restoration of each channel separately. Our approach performs the restoration of the \textit{YCbCr} channels in a single end-to-end model. 


\section{DL-based decoder for MPEG I-frame}
\label{sec:model}

Our proposal consists of an MPEG \textit{I-frame} image decoder that incorporates a DL model that learns how to recover lost coefficients in the encoding quantization process. Our method is purely based on the frequency-to-frequency domain. It reads the quantized DCT coefficients from a bitstream of low-quality \textit{I-frame} patch (QF=10) and predicts the missing coefficients to recompose the corresponding patch in high quality (QF=50). 

\begin{figure}[!ht]
    \centering
    \includegraphics[width=0.45\textwidth]{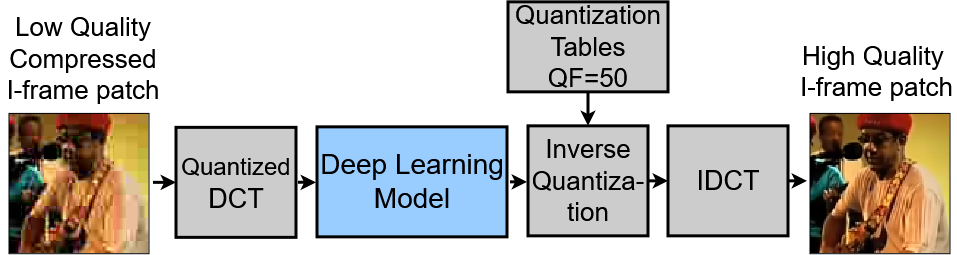}
    \caption{DL-based decoder for MPEG \textit{I-frame}.}
    \label{fig:model_flow}
\end{figure}

Fig.~\ref{fig:model_flow} illustrates the workflow of our method. Given a low-quality compressed \textit{I-frame} patch, the decoder first extracts the quantized DCT bitstream of \textit{YCbCr} channels and uses a DL model to restore these DCT coefficients. Then, it applies any of the default MPEG quantization tables (with a chosen QF of 50 in our example) for luminance and chrominance channels to the inverse quantization operation. Finally, it performs the inverse DCT (IDCT), level shifting, 8x8 block merging, and color space conversion from \textit{YCbCr} to \textit{RGB} to get the resulting high-quality \textit{I-frame} patch.

At the heart of our method lies the DL-Model, over which we have focused much of our efforts. In the remainder of this section, we detail the different DL networks we tested as possible backbones to our DL-based decoder. Among the many DL models found  in the literature, these were the ones that worked particularly well in our experiments (described in Section~\ref{sec:experiment}).

\subsection{DnCNN}

\begin{figure}[!ht]
    \centering
    \includegraphics[width=0.39\textwidth]{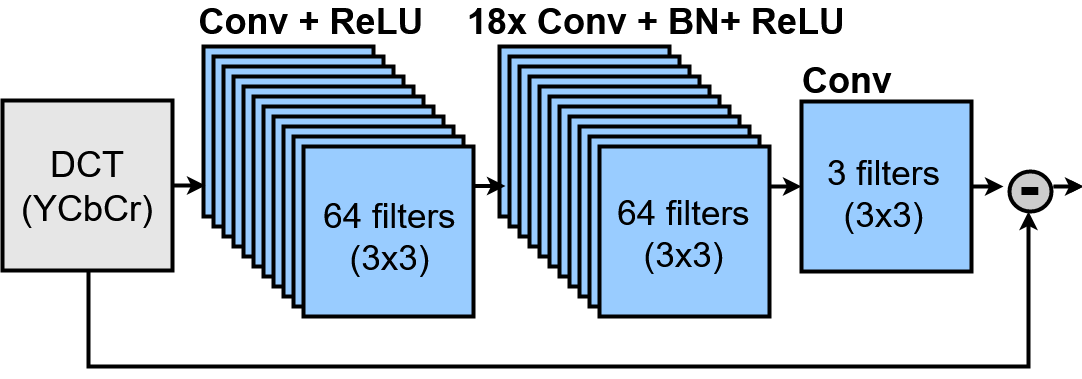}
    \caption{DnCNN architecture.}
    \label{fig:dncnn}
\end{figure}

 The DnCNN~\cite{zhang2017beyond} architecture is illustrated in Fig.~\ref{fig:dncnn}. This network is composed of three types of layers. The first layer, called \textit{Conv+ReLU}, consists of 64 filters with kernel size $3 \times 3$ and rectified linear units (ReLU) activation is utilized for nonlinearity. The second layer, called \textit{Conv+BN+ReLU}, added batch normalization~\cite{ioffe2015batch} between convolution and ReLu activation, this layer is repeated 18 times in sequence. The last layer uses 3 filters with kernel size $3 \times 3$ and linear activation to produce the residual features. Then, it is subtracted from the input by a global residual connection to reconstruct the output. 

\subsection{Deeper SRCNN}

\begin{figure}[!ht]
    \centering
    \includegraphics[width=0.39\textwidth]{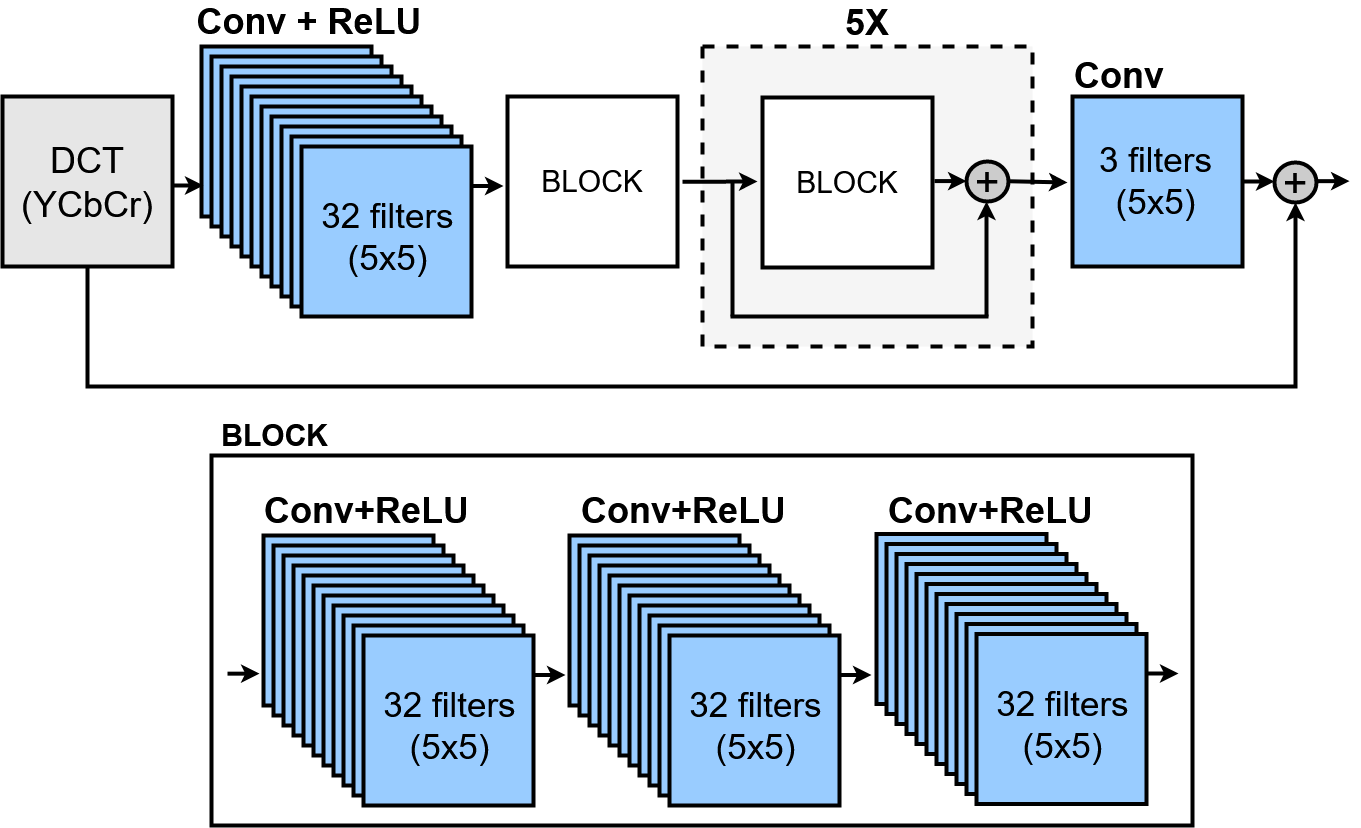}
    \caption{Deeper SRCNN architecture.}
    \label{fig:deeper_srcnn}
\end{figure}

Fig.~\ref{fig:deeper_srcnn} illustrates the Deeper SRCNN~\cite{arcnn2015} architecture. This network has 20 convolutional layers with 32 filters, kernel size $5 \times 5$, and ReLU activation, except for the last layer, which has 3 filters and linear activation. The block structure is composed of three convolutional layers with  ReLU activation. After the first block, the block structure is repeated five times, but with a residual connection interspersed between them. Then, the last layer is added to the input by a global residual connection to generate the output.

\subsection{ResNets}

\begin{figure}[!ht]
    \centering
    \includegraphics[width=0.39\textwidth]{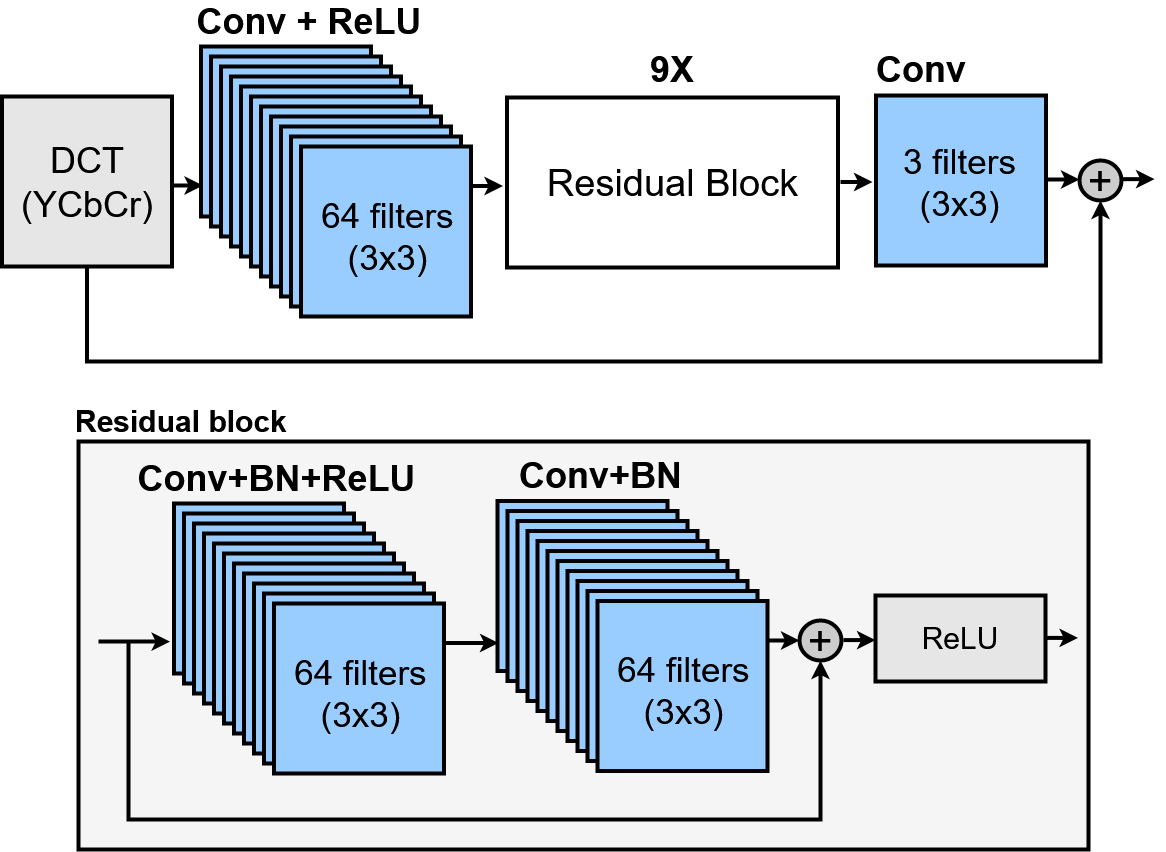}
    \caption{Resnet architecture with two residual block variants.}
    \label{fig:resnets}
\end{figure}

Fig.~\ref{fig:resnets} illustrates the ResNet network~\cite{he2016deep} adapted for image reconstruction.  It starts with a convolution layer with batch normalization and ReLU activation, followed by other convolution with batch normalization and linear activation. Next, a local residual connection is composed by the addition of the block input and second convolution output. Finally, a ReLU activation is performed to generate the block output. 

\subsection{Res-UNet}

\begin{figure}[!ht]
    \centering
    \includegraphics[width=0.45\textwidth]{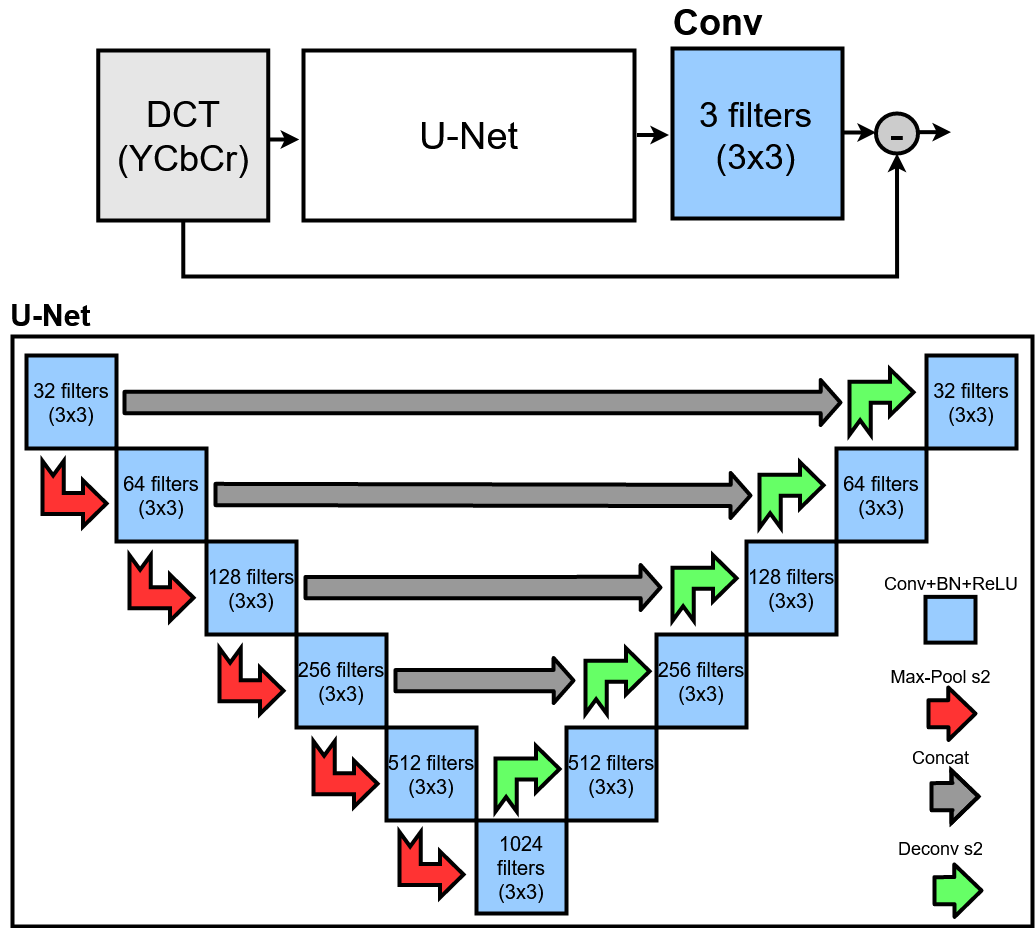}
    \caption{Res-UNet architecture.}
    \label{fig:runet}
\end{figure}

Fig.~\ref{fig:runet} illustrates the Res-UNet architecture, in which we have extended the plain U-Net~\cite{ronneberger2015u} with the addition of a global residual connection.  Global residual learning is present in all previous architectures. It is a trend in deep learning models for several image-to-image tasks, such as super-resolution, denoising, and artifacts removal~\cite{park2019densely, hou2019image, li2019fast}. 

The U-Net uses convolutional layers with batch normalization, kernel size 3x3, and ReLU activation. Our implementation uses 11 convolutional layers. The first five are downsampled by a max pooling with a kernel size of $3 \times 3$ and stride 2. Starting at the seventh convolution layer, before each convolutional layer, an upsampling is applied by a transpose-convolution with a kernel of size $3 \times 3$, stride 2, and concatenated with the output of the correspondent convolutional layer of the first half of U-Net.

\section{Experiment}
\label{sec:experiment}

Along this section, we evaluate the effectiveness of our models in restoring the quantized DCT coefficients. In order to attest the choice of our best model, we compare results obtained from the models presented in Section~\ref{sec:related} (DNCNN, Deeper SRCNN, ResNet, and Res-Unet). We evaluate our models using three popular metrics to measure the quality of reconstruction of lossy compression codecs: the PSNR (Peak Signal-to-Noise Ratio), the NRMSE (Normalized Root Mean Square Error), and the SSIM (Structural Similarity Index)~\cite{instruments2013peak, pinki2016estimation}.

As baseline models, we have chosen the AR-CNN and the Fast AR-CNN~\cite{arcnn2015} since they are simple models, and a few years ago, they were a reference for the JPEG artifact removal task in the spatial domain.\footnote{All the implementation code, including network models, training modules, saved models and datasets can be obtained form our public git repository at [omitted for double-blind-review].} 

The remainder of this Section is structured as follows. First, in Section~\ref{subsec:datasets}, we discuss the production of the two datasets, followed by Section~\ref{subsec:exp_setup}, where we describe our experimental setup. And, finally, our empirical findings and results are registered in Section~\ref{subsec:exp_result}.

\subsection{Datasets}
\label{subsec:datasets}

As listed in Table~\ref{tab:dataset}, we constructed a dataset based on patches extracted from 10 music videos from YouTube. We used the video ``Scorpions in Portugal" for training and validation, while the others just for testing. Using ffmpeg,\footnote{https://ffmpeg.org/} we extracted \textit{I-frames} at intervals of 1 frame-per-second. Next, we split each \textit{I-frame} in patches with common size of 120x120 pixels. 

\begin{table}[!ht]
    \centering
    \caption{Videos Dataset}
    \begin{tabular}{lcccc}
        \hline
        \textbf{Video Title} & \textbf{Frames} & \textbf{Train.} & \textbf{Valid.} & \textbf{Test} \\
        \hline
         Scorpions in Portugal & 216,197 & 95\% & 5\% & 0\% \\
         R. H. C. P. - Californication & 321 & 0\% & 0\% & 100\% \\
         M. J. - Smooth Criminal & 545 & 0\% & 0\% & 100\% \\
         Linkin Park - In The End  & 217 & 0\% & 0\% & 100\% \\
         Styx - Mr. Roboto & 334 & 0\% & 0\% & 100\% \\
         Z. Ramalho - Avôhai & 338  & 0\% & 0\% & 100\% \\
         Bee Gees - Stayin' Alive & 249 & 0\% & 0\% & 100\% \\
         R. Astley - N. G. G. You Up & 212 & 0\% & 0\% & 100\% \\
         Metallica - The Unforgiven & 649 & 0\% & 0\% & 100\% \\
         P.! At T. D. - High Hopes & 196 & 0\% & 0\% & 100\% \\
        \hline
    \end{tabular}
    \label{tab:dataset}
\end{table}

We implemented a MPEG-library to extract DCT coefficients from each \textit{I-frame} patch. Then, we quantized the DCT coefficients of each image using QF=10, QF=30, QF=40, and QF=50 and saved them as NumPy files. Samples with QF=10 are used as input for the models, while samples with QF=30, QF=40, and QF=50 are used as a reference during training and evaluation in three different experiments.

\subsection{Setup}
\label{subsec:exp_setup}

Our  networks  were  trained  using  an  octa-core  i7  3.40 GHz CPU   with   a   GTx-1070Ti   GPU.   The   training was based on the  Adam~\cite{kingma2014adam}  optimization  with  a  momentum  of  0.999,  an  exponential decay of 0.9,  and  epsilon of 1e-07,  a batch  normalization  with  a  decay of  0.9997  and  an epsilon  of  0.001 with a fixed  learning  rate  of  0.001, and MSE (Mean Square Error) as the loss function. We normalized both datasets and run our experiments for 400 epochs in all experiments.

\subsection{Results}
\label{subsec:exp_result}

In the first experiment, using QF=50 as reference (Table~\ref{tab:valid_exp1}), the best result was achieved by Res-UNet, which produced an SSIM of 88.05\%, PSNR of 33.62 and NRMSE of 12.92\%,  followed by ResNet, DnCNN, Deeper SRCNN, and U-Net. The other models had unsatisfactory performance since they produced worse results than the input data (MPEG QF=10). By analyzing the SSIM metric, we notice that the Res-UNet network produced an average improvement in the input data similar to the MPEG QF=20. This result shows that the Global Residual Learning is a useful technique since it improved the performance of Res-UNet when compared to basic U-Net.

\begin{figure*}
    \centering
    \includegraphics[width=\textwidth]{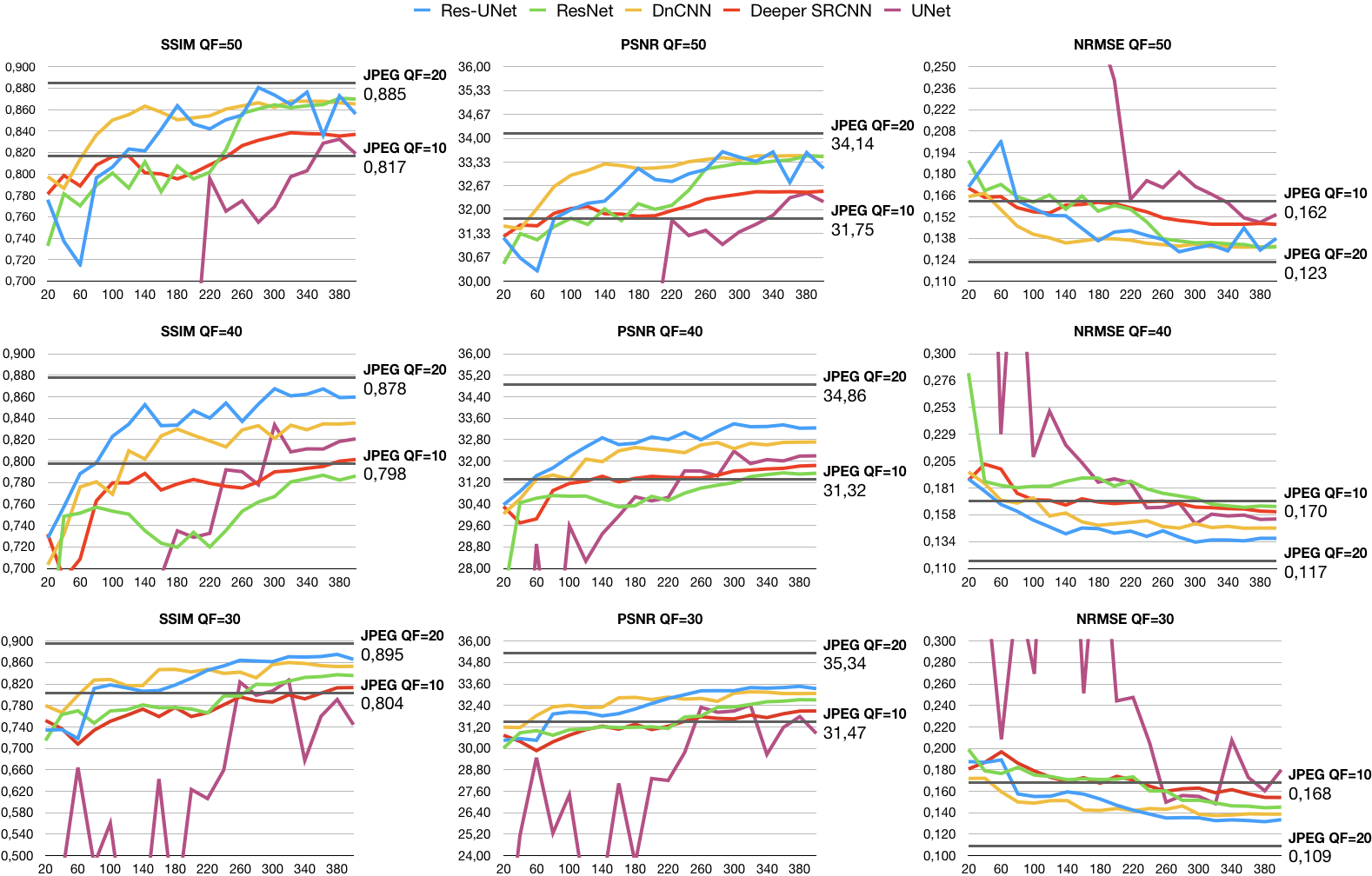}
    \caption{Convergence curve of Res-UNet, ResNet, DnCNN, Deeper SRCNN, and U-Net. The first, second and third row shows the average SSIM, PSNR and NRMSE on validation set for Dataset with QF=50, QF=40, and QF=30, respectively.}
    \label{fig:exp_chart}
    \vspace{-1em}
\end{figure*}

The first row of Fig.~\ref{fig:exp_chart} shows the convergence curve obtained during the training of the first experiment. The DnCNN network converged faster than the other networks; however, the Res-UNet network achieved its best result in epoch 280, improving the input data to slightly less than MPEG QF=20. Although the vanilla U-Net network managed to get closer to the other networks from the epoch 220, it produced the worst result at the end of the training.

The second and third experiments used QF=40 and QF=30 as references, respectively (Table~\ref{tab:valid_exp2} and Table~\ref{tab:valid_exp3}). The best result was also achieved by the Res-UNet in both experiments, which produced an SSIM of 86.74\%, PSNR of 33.39 in the second experiment, and an SSIM of 87.52\%, PSNR of 33.46 and NRMSE of 13.16\% in the third experiment. In comparison to the first experiment, all models produced worse results than those using DCT coefficients with QF=30 QF=40 as a reference. It is worth pointing out that the reference values for MPEG QF=20 change in each experiment because different reference QFs are used in the calculations.

The second and third rows of Fig.~\ref{fig:exp_chart} shows the convergence curve during the training of the second and third experiments respectively. The Res-UNet network converged faster and obtained the best performance in both scenarios. The other models had similar behavior to the ones produced in the first experiment, except for the ResNet network, which showed a considerable drop in performance.

\begin{table}[!ht]
    \centering
    \caption{Results of models on validation set with QF=50}
    \begin{tabular}{l l c c c }
        \hline
         \textbf{\#} & \textbf{Model} & \textbf{SSIM} & \textbf{PSNR} & \textbf{NRMSE} \\
        \rowcolor{lightgray}
        \hline
        \multicolumn{2}{c}{MPEG QF=20} & 0.884899 & 35.142163 & 0.113699 \\ 
        1 & Res-UNet & 0.880566 & 33.621739 & 0.129212 \\
        2 & ResNet & 0.870629 & 33.504199 & 0.132054 \\
        3 & DnCNN & 0.867987 & 33.510900 & 0.132198 \\
        4 & Deeper SRCNN & 0.838423 & 32.517377 & 0.147069 \\
        5 & U-Net & 0.832437 & 32.461556 & 0.148408 \\ 
        \rowcolor{lightgray}
        \multicolumn{2}{c}{MPEG QF=10 (input)} & 0.816729 & 31.755941 & 0.162004\\
        6 & AR-CNN & 0.775617 & 31.170242 & 0.171788 \\
        7 & Fast AR-CNN & 0.774667 & 31.298347 & 0.170258 \\
        \hline
    \end{tabular}
    \label{tab:valid_exp1}
\end{table}

\begin{table}[!ht]
    \centering
    \caption{Results of models on validation set with QF=40}
    \begin{tabular}{l l c c c }
        \hline
        \textbf{\#} & \textbf{Model} & \textbf{SSIM} & \textbf{PSNR} & \textbf{NRMSE} \\
        \rowcolor{lightgray}
        \hline
        \multicolumn{2}{c}{MPEG QF=20} & 0.877939 & 34.869542 & 0.116706 \\
        1 & Res-Unet & 0.867491 & 33.392234 & 0.133276 \\
        2 & DnCNN & 0.835562 & 32.715130 & 0.145711 \\
        3 & U-Net & 0.833832 & 32.373612 & 0.149488 \\
        4 & Deeper SRCNN & 0.801505 & 31.839737 & 0.160213 \\
        \rowcolor{lightgray}
        \multicolumn{2}{c}{MPEG QF=10 (input)} & 0.797985 & 31.323203 & 0.169981 \\ 
        5 & ResNet & 0.786911 & 31.563261 & 0.164136 \\
        6 & AR-CNN & 0.760004 & 31.005140 & 0.176942 \\
        7 & Fast AR-CNN & 0.752536 & 30.672298 & 0.181181 \\
        \hline
    \end{tabular}
    \label{tab:valid_exp2}
\end{table}

\begin{table}[!ht]
    \centering
    \caption{Results of models on validation set with QF=30}
    \begin{tabular}{l l c c c }
        \hline
        \textbf{\#} & \textbf{Model} & \textbf{SSIM} & \textbf{PSNR} & \textbf{NRMSE} \\
        \rowcolor{lightgray}
        \hline
        \multicolumn{2}{c}{MPEG QF=20} & 0.895385 & 35.342902 & 0.109367 \\ 
        1 & Res-Unet & 0.875200 & 33.466494 & 0.131608 \\
        2 & DnCNN & 0.860004 & 33.161343 & 0.138657 \\
        3 & ResNet & 0.837271 & 32.721260 & 0.144566 \\
        4 & U-Net & 0.827067 & 32.423205 & 0.148407 \\ 
        5 & Deeper SRCNN & 0.813350 & 32.092821 & 0.154192 \\ 
        \rowcolor{lightgray}
        \multicolumn{2}{c}{MPEG QF=10 (input)} & 0.803636 & 31.470544 & 0.168134 \\ 
        6 & Fast AR-CNN & 0.778225 & 31.193668 & 0.170051 \\
        7 & AR-CNN & 0.747525 & 30.395390 & 0.185941 \\
        \hline
    \end{tabular}
    \label{tab:valid_exp3}
\end{table}

\begin{figure}[ht]
    \centering
    \includegraphics[width=0.46\textwidth]{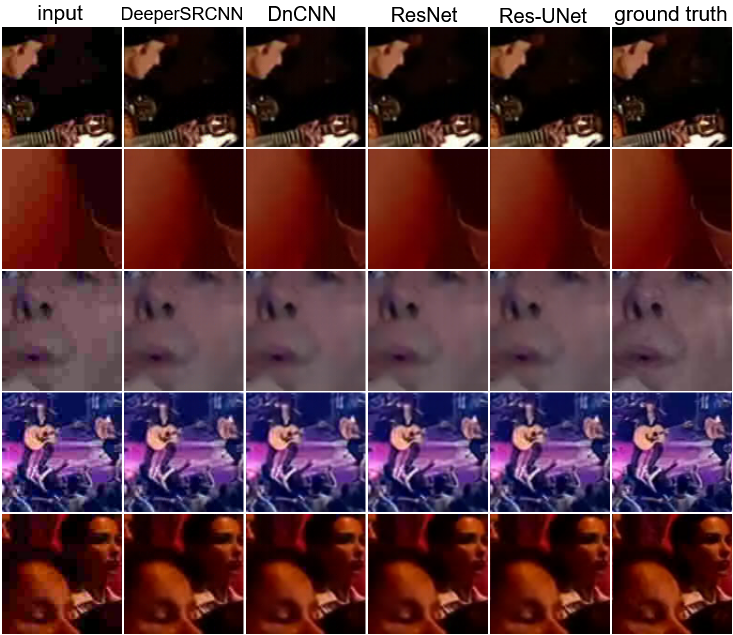}
   \caption{I-frame patches restored by DCT coefficients prediction made by the DeeperSRCNN, DnCNN, ResNet, and Res-UNet networks on the validation set.}
    \label{fig:valid_frames_img}
    \vspace{-1em}
\end{figure}

Fig.~\ref{fig:valid_frames_img} shows a comparison of four prediction examples produced by Deeper SRCNN, DnCNN, ResNet, Res-Unet, and DnCNN networks. Notice that the DCT coefficients restored by the Res-UNet network produced an image without block structures and recovered some texture details of the original image. However, although the resulting image is better than the input image, it still has a blurred aspect when compared to the ground-truth image.

The results of the best model with the test videos are summarized in Table~\ref{tab:table_test}. Our proposed decoder was able to improve the visual quality in most of the videos, except for the ``Metallica - The Unforgiven'' video, which produced a lower than expected result. We checked the content of the video and realized that this particular video contains DCT frequencies that are rare in the training set. This issue can be solved by adding more data with these characteristics to the training set.

\begin{figure*}
    \centering
    \includegraphics[width=0.9\textwidth]{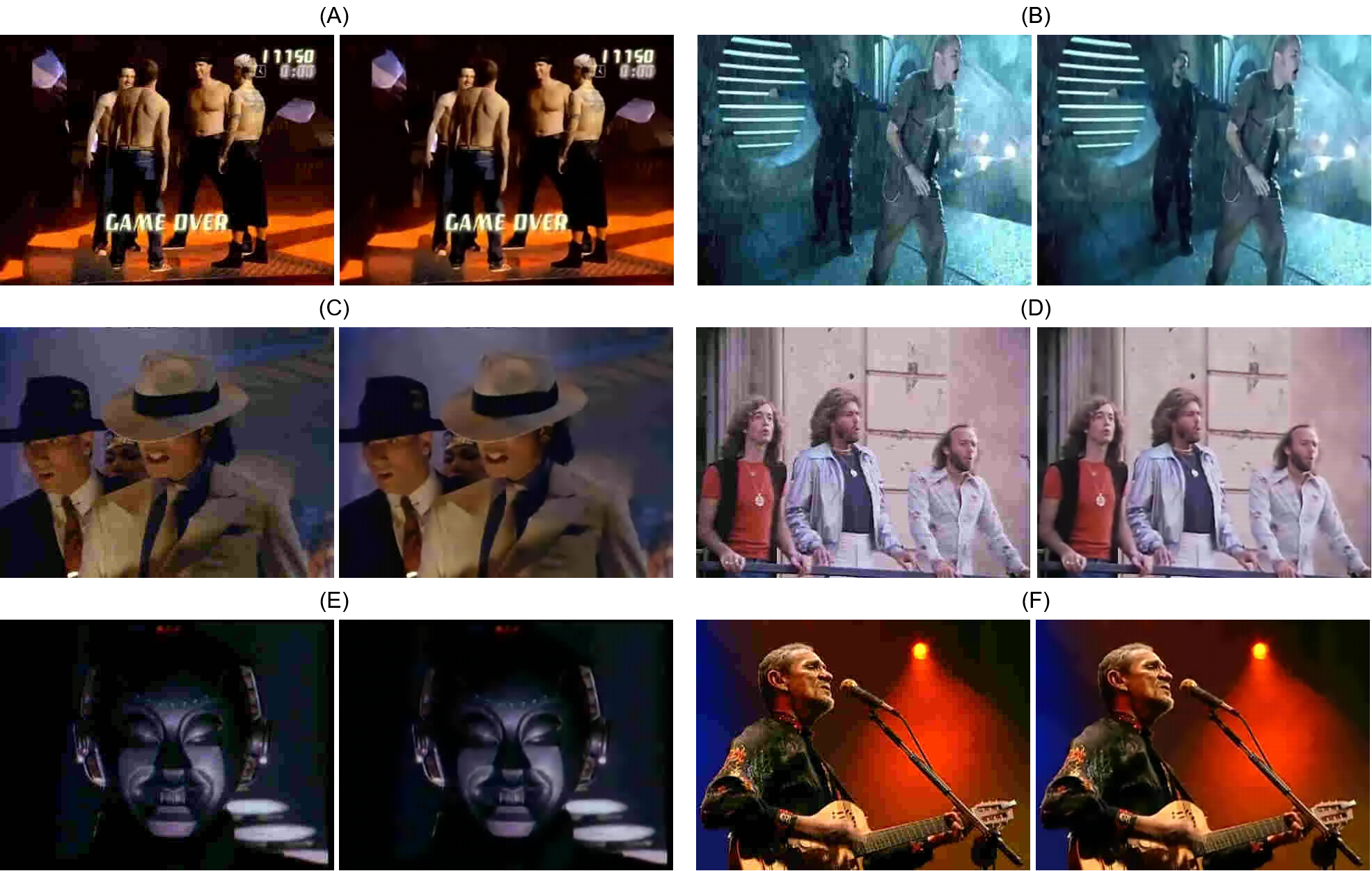}
    \caption{Video frames of the test set restored by DCT coefficients prediction produced by Res-UNet. Videos: (A) R. H. C. P. - Californication; (B) Linkin Park - In The End; (C) M. J. - Smooth Criminal; (D) Bee Gees - Stayin’ Alive; (E) Styx - Mr. Roboto; And (F) Z. Ramalho - Avôhai} 
    \label{fig:best_test}
     \vspace{-1em}
\end{figure*}

Fig.~\ref{fig:best_test} shows some frames of the test videos restored by DCT coefficients prediction produced by Res-UNet.\footnote{The full demonstration of our model is available on Youtube: Youtube link [Removed for double-blind-review].}
It is worth mentioning that although the visual quality has improved a lot, there are still some continuity faults between the frame patches.

\begin{table}[!ht]
    \centering
    \caption{Results of Res-UNet on test videos}
    \begin{tabular}{lllll}
        \hline
        \textbf{Video Title} & \textbf{} & \textbf{SSIM} & \textbf{PSNR} & \textbf{NRMSE} \\
        \hline
         \tiny{R. H. C. P.} & QF=10 & 0.8497 & 30.62 & 0.1534 \\
         Californication & Res-Unet & 0.8738 & 30.71 & 0.1227 \\ \hline
          \tiny{M. J.} & QF=10 & 0.8320 & 31.60 & 0.1066 \\
         Smooth Criminal & Res-Unet & 0.8769 & 32.77 & 0.1531 \\ \hline
          \tiny{Linkin Park}  & QF=10 & 0.8156 & 31.52 & 0.1740 \\
         In The End & Res-Unet & 0.8764 & 32.78 & 0.1738 \\ \hline
          \tiny{Styx} & QF=10 & 0.8318 & 33.50 & 0.2562 \\
         Mr. Roboto & Res-Unet & 0.8790 & 34.37 & 0.1887 \\ \hline
          \tiny{Z. Ramalho} & QF=10 & 0.7730 & 32.17 & 0.2255 \\
         Avôhai & Res-Unet & 0.8698 & 32.18 & 0.1876 \\ \hline
          \tiny{Bee Gees} & QF=10 & 0.8467 & 31.20 & 0.0756 \\
         Stayin' Alive & Res-Unet & 0.8876 & 31.17 & 0.0747 \\ \hline
          \tiny{R. Astley} & QF=10 & 0.8334 & 32.19 & 0.1479 \\
         N. G. G. You Up & Res-Unet & 0.8764 & 32.27 &  0.0839 \\ \hline
          \tiny{Metallica} & QF=10 & 0.8639 & 31.18 & 0.2023 \\
        The Unforgiven & Res-Unet & 0.8456 & 30.93 & 0.2271 \\ \hline
          \tiny{P.! At T. Disco} & QF=10 & 0.8469 & 31.27 & 0.1876 \\
        High Hopes & Res-Unet & 0.8698 & 32.17 & 0.1003 \\ 
        \hline
    \end{tabular}
    \label{tab:table_test}
  
\end{table}

\section{Conclusion}
\label{sec:conclusion}

In this work we have proposed a MPEG \textit{I-frame} decoder that incorporates a DL model that learns how to recover the lost coefficients in the MPEG-encoding quantization process. We have made our experiments with our produced video datasets in order to select the best model based on the SSIM, PSNR, and NRMSE metrics. Among the seven DL analyzed networks, the Res-UNEt was the one that achieved the best performance in both datasets. We have empirically shown that, given the DCT coefficients with QF=10, our decoder using DnCNN can produce DCT coefficients similar to the QF=20 quality --- when using DCT coefficients with QF=50 as reference.

As a future work, we intend to apply some of the current trends coming from the deep learning field (e.g., spatial attention~\cite{ye2016spatial}, channel-wise attention~\cite{zhu2019stacked}, and transformer encoding~\cite{kasem2018revised}) to extend our model. We want to test whether these techniques are useful for the DCT coefficients restoration task.

We also plan to investigate if we can improve the video quality by taking advantage of the MPEG-2 standard~\cite{watkinson2004mpeg}. Another interesting investigation would be the extension of our decoder to include enhancements to the quality of \textit{P/B-frames}. We also noticed a visual continuity gap between the frame patches in the experiment; we intend to address this problem with an extension that smooths the connections between frame patches.

\bibliographystyle{IEEEtran}
\bibliography{IEEEabrv,bare_conf.bib}
%

\end{document}